\documentclass{acm_proc_article-sp}

\begin{document}

\title{Augmented Test Collections: A Step in the Right Direction}
%
%
%
%
%

\numberofauthors{3} 
%
\author{
%
%
\alignauthor
Laura Hasler\\
       \affaddr{Information Retrieval Group}\\
       \affaddr{Information School}\\
       \affaddr{University of Sheffield}\\
       \email{l.j.hasler@sheffield.ac.uk}
\alignauthor
Martin Halvey\\
       \affaddr{ITT}\\
       \affaddr{Engineering and Built Environment}\\
       \affaddr{Glasgow Caledonian University}\\
       \email{martin.halvey@gcu.ac.uk}
\alignauthor Robert Villa\\
       \affaddr{Information Retrieval Group}\\
       \affaddr{Information School}\\
       \affaddr{University of Sheffield}\\
       \email{r.villa@sheffield.ac.uk}
}
\additionalauthors{Additional authors: John Smith (The Th{\o}rv{\"a}ld Group,
email: {\texttt{jsmith@affiliation.org}}) and Julius P.~Kumquat
(The Kumquat Consortium, email: {\texttt{jpkumquat@consortium.net}}).}
\date{30 July 1999}

\toappear{\the\boilerplate\par
{\confname{\the\conf}} \the\confinfo\par \the\copyrightetc}

\permission{Copyright is held by the author/owner(s).}
\conferenceinfo{SIGIR'14 Workshop on Gathering Efficient Assessments of Relevance (GEAR'14),}{\\July 11, 2014, Gold Coast, Queensland, Australia.}
\copyrightetc{}

\maketitle

\begin{abstract} 
In this position paper we argue that certain aspects of relevance assessment in the evaluation of IR systems are oversimplified and that human assessments represented by qrels should be augmented to take account of contextual factors and the subjectivity of the task at hand. We propose enhancing test collections used in evaluation with information related to human assessors and their interpretation of the task. Such augmented collections would provide a more realistic and user-focused evaluation, enabling us to better understand the evaluation process, the performance of systems and user interactions. A first step is to conduct user studies to examine in more detail what people actually do when we ask them to judge the relevance of a document.
\end{abstract}




\section{Introduction}

Training and test collections are a fundamental part of the process of developing effective retrieval systems, and are used extensively in evaluation forums such as TREC and CLEF. These collections involve human input to assess the relevance of documents to a given topic, and the assessments are considered a benchmark by which we train and evaluate our systems. However, in the resulting collections, the human process of judging relevance is oversimplified. It may take substantial effort for humans to judge the relevance of documents, where numerous factors come into play and context is crucial. The level of subjectivity of the task can be high and agreement on ''relevance'' low, yet the information used to evaluate our retrieval systems is encapsulated in a single number in a qrel and is taken as a ''gold standard''. In this paper, we argue that to move forward and advance our field, we need to acknowledge that we oversimplify things and that the standard collections with which we evaluate our systems are fairly synthetic. We need to be prepared to take on the challenge of considering the less well-defined and perhaps messy aspects of information retrieval evaluation.

Given that humans are necessarily involved in the process of developing test collections, and that the process of relevance assessment upon which these collections depend is not as simple as the collections suggest, a good starting point is to examine in more detail what people actually do when we ask them to judge the relevance of documents to a particular topic or information need. We need to do this across different types of document, different types of task, and different types of assessor. By exploring what it is that people really do and by considering the assessors themselves, we can augment the final relevance judgements with additional contextual information about the assessor and their interpretation of the task. This additional information could help to explain subjective and complex relevance judgements in their particular contexts and reflect how users of systems might also consider results to be relevant, capturing a range of perspectives on the relevance of documents. Augmented test collections containing this type of information would therefore allow a more realistic evaluation of retrieval engines from a user perspective, enabling us to better understand both their performance and user interactions.

\section{Augmented Test Collections}

The complexity of the relevance assessment task has long been recognised in the literature, with a focus on relevance criteria ~\cite{barry1994user,barry1998users,saracevic1970concept,xu2006relevance}. More recently, a number of studies have explored factors affecting the process of performing relevance assessments and search tasks, considering aspects such as level of expertise or types of judges ~\cite{kinney2008evaluator,clough2013examining,webber2013assessor}, task complexity ~\cite{bell2004searcher}, certainty ~\cite{aluser} and effort ~\cite{villa2013relevance} for different types of documents. However, the valuable information resulting from such studies has not found its way into the practical side of system evaluation - it is not present in test collections themselves.

To enable a more realistic and user-oriented evaluation in IR, we propose the development of augmented test collections. In these collections, anonymised information including standard demographic information such as age, gender and level of education, along with task-specific information relating to assessor expertise, interest, motivation, confidence or certainty, degree of relevance of a document and reasons that a document is considered relevant is added for each relevance assessment. To achieve this, we first need to investigate the factors which affect the judging process, taking into account different judges, document types and tasks. We need to be able to ask the judges about their decisions to ensure that we fully and accurately understand the process and their interpretation of it.  

We advocate small-scale, qualitative lab-based studies of assessors as the best method to begin our exploration. Our studies will take into account much more than ''is this document relevant to this topic?'', and as well as performing assessments on standard collections, participants will be asked to specify their own information needs and assess documents as relevant to those. This is an attempt to create a more realistic approximation of real-life relevance assessment in contrast to the secondary assessments made on synthetic benchmarks and will be informed by user-centred library and information science methods (e.g. put back in reference). We intend to use a range of document types (text, image and video) and a range of assessors to obtain more comprehensive insights. We think that some of the most valuable insights from our studies will be gained through playing back assessors judgements to them, asking them to talk through the process and asking why they made certain decisions, then analysing this discussion to establish themes linked to relevance judgements. Assessors will also answer predefined questions regarding their levels of expertise, interest and motivation for a topic, and their perception of the difficulty of the task and the effort expended. Whilst this is a time- and labour-intensive method, it provides rich insights that would be difficult to obtain through other methods. The information collected would provide the basis for the augmented test collections we propose, where it would be attached to its corresponding qrel. This additional information will help to make the hidden elements of subjectivity and complexity in relevance assessments within test collections more transparent, as well as to help explain the more measurable differences in assessor judgements (i.e. as captured by inter-assessor agreement metrics). 

A next step would be to investigate whether we can gather a substantial amount of comparable assessments and corresponding additional information effectively on a larger scale, using, for example, crowdsourcing as a method of data collection. Whilst crowdsourcing experiments cannot generate the same degree of qualitative data as the lab setting, the number of participants in a single experiment will be much greater, as will the resulting quantity of assessments generated, enabling us to explicitly test the factors and themes identified in the lab on a larger scale. These studies will also allow us to investigate whether it is feasible to create our proposed type of augmented test collection using a less qualitative methodology. Considering the amount of time, effort and money it takes to produce test collections and that often large numbers of assessments are needed for evaluation, as well as the problem of subjectivity, it is sensible to explore alternative means of gathering judgements.

\section{Conclusions}
In this position paper, we acknowledged the oversimplification of the human process of relevance assessment as 
represented in IR evaluation, which contributes to the opaque nature of the benchmarks we use. We proposed the development of augmented test collections, where qrels are enhanced with additional information regarding the assessor and their interpretation of the task, as a way forward. The contextual information used to augment the qrels would help to make the subjectivity and complexity associated with relevance assessments more explicit within test collections, leading to a potentially more natural and user-focused approach to the evaluation of systems which are, after all, used by real people who make these judgements during their everyday interactions. We outlined a possible way of exploring the factors necessary for the development of such collections through qualitative lab-based studies which can then be scaled up via crowdsourcing. We believe that our proposal of augmented test collections is a step in the right direction towards more realistic, clear and useful IR evaluation. 

\section{Acknowledgements}
This work was funded by the UK Arts and Humanities Research Council (grant AH/L010364/1).

%
\bibliographystyle{abbrv}
\bibliography{gear2014-hasler}  

\end{document}